\documentclass[prb,aps,twocolumn,showpacs,superscriptaddress,floatfix]{revtex4}
\usepackage{epsfig}
\usepackage{amsmath}
\usepackage{amsfonts}
\usepackage{mathptmx}
\usepackage{eucal}
\usepackage{bm}

\DeclareMathOperator{\Tr}{\mathop{\mathrm{Tr}}}

\DeclareMathOperator{\re}{\mathop{\mathrm{Re}}}
\DeclareMathOperator{\im}{\mathop{\mathrm{Im}}}
\DeclareMathOperator{\arctanh}{arctanh}

\newcommand{\Eq}[1]{Eq.~(\ref{#1})}
\newcommand{\Eqs}[1]{Eqs.~(\ref{#1})}

\begin{document}

\title{Subgap current in superconducting tunnel junctions with diffusive electrodes}

\author{E.~V.~Bezuglyi}
\affiliation{Institute for Low Temperature Physics and
Engineering, Kharkov 61103, Ukraine}
\affiliation{Chalmers University of Technology, S-41296
G\"oteborg, Sweden}

\author{A.~S.~Vasenko}
\affiliation{Chalmers University of Technology, S-41296
G\"oteborg, Sweden}
\affiliation{Department of Physics, Moscow State University,
Moscow 119992, Russia}

\author{E.~N.~Bratus'}
\affiliation{Institute for Low Temperature Physics and
Engineering, Kharkov 61103, Ukraine}

\author{V.~S.~Shumeiko}
\affiliation{Chalmers University of Technology, S-41296
G\"oteborg, Sweden}

\author{G.~Wendin}
\affiliation{Chalmers University of Technology, S-41296
G\"oteborg, Sweden}

\date{\today}

\begin{abstract}
We calculate the subgap current in planar superconducting tunnel
junctions with thin-film diffusive leads. It is found that the
subharmonic gap structure of the tunnel current scales with an
effective tunneling transparency which may exceed the junction
transparency by up to two orders of magnitude depending on the
junction geometry and the ratio between the coherence length and
the elastic scattering length. These results provide an
alternative explanation of enhanced values of the subgap current
in tunneling experiments often ascribed to imperfection of the
insulating layer. We also discuss the effect of finite lifetime of
quasiparticles as the possible origin of additional enhancement of
multiparticle tunnel currents.
\end{abstract}

\pacs{74.45.+c, 74.40.+k, 74.25.Fy, 74.50.+r}

\maketitle

Subgap quasiparticle current in superconducting junctions at small
applied voltages $eV<2\Delta$ is the subject of persistent
theoretical interest and experimental research. Recently, the
problem has attracted new attention, and a number of measurements
of the subgap current in high-quality tunnel junctions have been
performed,\cite{Gubrun2001,Lang2003} motivated by the problem of
decoherence in Jo\-seph\-son-junction-based superconducting
qubits.\cite{Makhlin} The subgap current at zero temperature is
due to multiparticle tunneling (MPT) processes,\cite{MPT} whose
intensities strongly depend on the quality of the insulating
layer, being enhanced by disorder, localized electronic states,
pinholes, etc.\cite{KBT} The effect of disorder in the junction
electrodes on the subgap current has never been questioned.

According to the MPT theory,\cite{MPT} the subgap tunnel current
depends on the transparency $D$ of the tunnel barrier: it
decreases with decreasing voltage in a steplike fashion with step
heights proportional to $(D/2)^n$ at voltages $eV=2\Delta/n$,
$n=1,2...$ [subharmonic gap structure (SGS)]. Similar results have
been obtained for junctions with ballistic
electrodes,\cite{Bratus95} and mesoscopic point contacts with
diffusive electrodes\cite{Scheer} on the basis of the theory of
multiple Andreev reflections (MAR).\cite{KBT}

Experimentally, the SGS scaling parameter in atomic size junctions
nicely agrees with the theory;\cite{Jan2000} however, in
macroscopic tunnel junctions it is usually much larger
\cite{Gubrun2001,Lang2003} (see also earlier data
\cite{Cristiano}); moreover, there is a smooth residual current at
a very low voltage.\cite{Gubrun2001} Although enhanced SGS in
high-trans\-mission junctions could be explained by assuming
randomly distributed resonant levels within the tunnel
barrier,\cite{Likharev} enhanced subgap current in
low-transmission junctions with presumably good insulating layers
remains an open question.

\begin{figure}[bt]
\epsfxsize=8.5cm\epsffile{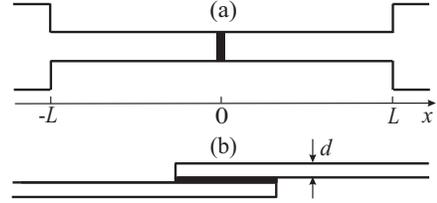} \vspace{-2mm}
\caption{One-dimensional (a) and planar (b) models of the tunnel
junction with diffusive leads; equilibrium reservoirs are far from
the contact, $L \gg\xi_0$.} \label{model} \vspace{-4mm}
\end{figure}

In this paper we reexamine the problem of the subgap current in
{\em macroscopic} tunnel junctions, and consider the effects of
diffusive electrodes and planar junction geometry common for the
experiment (see Fig.~\ref{model}). Our main result is that the SGS
scaling parameter for such junctions significantly exceeds the
junction transparency: for the sandwich-type junction with
thin-film leads shown in Fig.~\ref{model}(b), the scaling is
determined by the effective transparency defined as
\begin{equation}\label{Deff}
D_{\textit{eff}} = (3\xi_0^2/\ell d)D,
\end{equation}
where $\xi_0=\sqrt{\mathcal{D}/2\Delta}$ is the diffusive
coherence length (we assume $\hbar = k_B =1)$, $\ell$ is the
elastic mean free path, $d \ll \xi_0$ is the thickness of the
leads, and $\mathcal{D}$ is the diffusion coefficient. For Al
junctions with $\ell \sim d = 50$ nm and $\xi_0 = 300$ nm, the
enhancement factor approaches 100. For the junctions with
one-dimensional (1D) geometry of Fig.~\ref{model}(a),
$D_{\textit{eff}} = (3\xi_0/\ell)D$.\cite{Jos} This result also
applies to nonhomogeneous tunnel barriers as soon as the size of
pinholes (more transparent spots) exceeds the elastic mean free
path, otherwise the ballistic scaling\cite{Bratus95} will be
valid.

The enhancement effect can be qualitatively understood by
considering a short point contact with the reservoirs located very
close to the contact, $L\ll\xi_0$ [cf. Fig.~\ref{model}(a)]. In
this case, the current can be calculated within the mesoscopic
approach,\cite{Averin97} by integrating over contributions of
normal conducting eigenmodes with randomly distributed
transparencies. The relevant distribution is known to be spread
over the interval $\sim (L/\ell)D \gg D$.\cite{Beenakker} The most
transparent modes dominate the subgap current, giving
$D_{\textit{eff}} \sim (L/\ell)D$. In our case of junctions with
large distance to the reservoirs, the scale of the spatial
variation of the Green's function $\xi_0$ plays the role of the
effective junction length giving qualitatively our result,
$D_{\textit{eff}} \sim (\xi_0/\ell)D$.\cite{note} We note that for
the long junctions under consideration the statistics of the
eigenmode transparencies is not known, and a quantitative result
has to be derived from the quasiclassical theory for diffusive
superconductors.

Our analysis is based on the diffusive equations \cite{LOnoneq}
for the quasiclassical two-time Keldysh-Green functions
$\check{G}(\bm{r},t_1,t_2)$,
\begin{align} \label{Keldysh}
[\check{H},\circ\check{G}]=i {\mathcal D}\nabla \check{J}, \quad
\check{G}\circ \check{G}=1,\quad
\quad \check{G} =\begin{pmatrix}
\hat{g}^R & \hat{G}^K \\
0 & \hat{g}^A \end{pmatrix}.
\end{align}
Here $\check{H}(t_1,t_2)=[i\sigma_z\partial_{t_1} + \Delta
\exp(i\sigma_z \phi)i\sigma_y]\delta(t_1-t_2)$, $\phi$ is the
superconducting phase, the sign $\circ$ denotes time convolution,
and $\check{J}=\check{G} \circ \nabla \check{G}$. Equation
\eqref{Keldysh} can be decomposed into the Us\-a\-del equation for
the retar\-ded or advan\-ced Green's functions $\hat{g}^{R,A}$ and
the equation for the Keldysh function $\hat{G}^K = \hat{g}^R
\circ\hat{f} - \hat{f}\circ \hat{g}^A$, where $\hat{f} = f +
\sigma_z f_-$ is the distribution function.

We present detailed calculations for the simpler, 1D geometry of
Fig.~\ref{model}(a). At the left electrode, $x = -L$, the Fourier
transformations of the two-time functions $\hat{g}$ and $\hat{f}$
with respect to the variable $t_1-t_2$ are given by the
equilibrium expressions
\begin{align}
\hat{g}(E) &= \sigma_z u(E) + i\sigma_y v(E), \quad
\hat{f}(E)=\tanh({E}/{2T}), \label{g0}\\
(u,v)&=(E,\Delta)/\xi(E), \quad \xi^{R,A} = [{(E\pm
i0)^2-\Delta^2}]^{1/2}.
\end{align}
At the right voltage-biased electrode, $x=L$, the function
$\check{G}$ is defined through the gauge
transformation\cite{Artemenko}
\begin{align} \label{gauge}
\check{G}(L) = \overline{\check{G}}(-L) \equiv S(t_1)
\check{G}(-L) S^\dagger(t_2),
\end{align}
with a unitary operator $S(t) =\exp[i\sigma_z\phi(t)/2]$, where
the phase $\phi$ satisfies the Josephson relation $\phi(t) =
2eVt$.

The boundary conditions \cite{KL} for the functions $\check{G}$
and $\check{J}$ at the tunnel barrier ($x=\pm 0$) are given by the
relations
\begin{equation}
\check{J}_{-0} = \check{J}_{+0} = \frac{W}{\xi_0}
\bigl[\check{G}_{-0},\circ \check{G}_{+0}\bigr],\quad W=
\frac{R_0}{2R} =\frac{3\xi_0}{4\ell}D,\label{Boundary2}
\end{equation}
where $R$ is the resistance of the tunnel barrier, $R_0= \xi_0/g$
is the resistance of a piece of the lead with length $\xi_0$, and
$g$ is the conductance of the leads per unit length. Assuming a
small value of the tunneling parameter $W$, we neglect the charge
imbalance function $f_-$ and the superfluid momentum within the
leads, as well as the variation of $\Delta$. In such an
approximation, \Eq{gauge} extends to the whole right lead,
$\check{G}(x) = \overline{\check{G}}(-x)$ for $0 < x < L$. The
problem is therefore reduced to the solution of a static equation
for the function $\check{G}(x,t_1,t_2)$ at $-L<x<0$ with the
time-dependent boundary condition \eqref{Boundary2} at $x=-0$. The
electric current is related to the Keldysh component $\hat{J}^K$
of the matrix current $\check{J}$ as $I(t)=({\pi g}/{4e}) \Tr
\sigma_z \hat{J}^K(x,t,t)$.\cite{LOnoneq} Using \Eqs{gauge} and
\eqref{Boundary2}, it can be expressed as
\begin{equation}
I(t)=(\pi /8eR) \Tr \sigma_z
[\check{G},\circ\overline{\check{G}}]^K(t,t). \label{Curr2}
\end{equation}
In this and following equations, the functions are taken at the
boundary $x=-0$. Expanding all functions over harmonics of the
Josephson frequency, $A(E,t)=\sum\nolimits_m A(E,m)e^{-2ieVm t}$
[$t=(t_1+t_2)/2$], we arrive at the equation for the dc current
$I$,
\begin{align}
I &=  \frac{1}{16eR}\int_{-\infty }^\infty dE\,  \Tr
\sum\nolimits_m \bigl[\hat{h}(E,m)\overline{\hat{G}^K}(E,-m)\nonumber
\\
&-\overline{\hat{h}}(E,m) \hat{G}^K(E,-m)\bigr], \quad \hat{h}=
\sigma_z \hat{g}^R - \hat{g}^A \sigma_z. \label{I0}
\end{align}

In the tunneling limit $W\ll 1$, the amplitude of the $m$th
harmonic is proportional to $W^m$; thus the zero harmonic $m=0$ of
the functions $\hat{g}$ and $\hat{G}^K$ in \Eq{I0} plays the key
role, while the high-order harmonics can be neglected. The effect
of these harmonics will be discussed later. Within this
approximation, the Green's function matrix structure in \Eq{g0}
holds, and the current \Eq{I0} {\em exactly} transforms to the
form
\begin{equation}\label{I00}
I = \frac{1}{eR}\int_{-\infty }^\infty dE\, N(E) N(E-eV)
[n(E-eV)-n(E)].
\end{equation}
Here $N(E)=\re u^R$ is the density of states (DOS) normalized to
its value in the normal state, and the distribution function $n=
(1/2)(1-f)$ approaches the Fermi function $n_F$ in equilibrium.
Furthermore, we split the integral in \Eq{I00} into  pieces of
length $eV$, denoting $A_k(E) \equiv A(E+keV)$,
\begin{align}\label{I001}
I &= \frac{1}{eR}\int_{0}^{eV}dE \, J(E), \quad J(E)=
\sum\nolimits_{k=-\infty}^\infty j_k(E), \\
j_k &= ({n_{k-1} - n_k}){\rho_k^{-1}}, \quad \rho_k^{-1}= N_k
N_{k-1}.\label{I002}
\end{align}
The distribution function $n(E,x=-0)$ satisfies the recurrence
relation following from the kinetic equation
$\partial_x\left(D_+\partial_x n\right) = 0$,
\begin{equation}\label{recurr}
\Theta(|E_k|-\Delta)[n_F(E_k) -n_k] = r (j_{k+1} - j_k),
\end{equation}
where $\Theta(x)$ is the Heaviside step function, $r = R_N/R \ll
1$, and $R_N$ is the normal resistance of the lead. To justify
\Eq{recurr}, we note that the diffusion coefficient $D_+= (1/2)(1
+ |u|^2 - |v|^2)$ is approximately constant, $D_+ \approx 1$, at
$|E|>\Delta$, which leads to the linear function $n(E,x)=n_{-0} +
(x/L)(n_{-0}-n_F)$. At $|E|<\Delta$, $D_+$ turns to zero at $|x|
\gg \xi_0$, which reflects complete Andreev reflection in the
leads and results in zero probability current, $D_+\partial_x n
=0$. Then, using the boundary condition at the tunnel barrier
following from \Eq{Boundary2}, $D_+\partial_x n =
(2W/\xi_0)\sum\nolimits_{k=\pm 1}NN_k (n_k - n)$, we arrive at
\Eq{recurr}.

\begin{figure}[tb]
\epsfxsize=8.5cm\epsffile{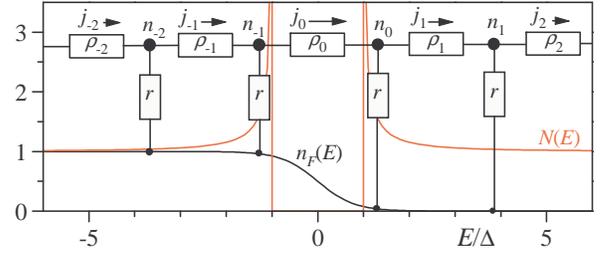}\vspace{-3mm}
\caption{Circuit representation of charge transport in a diffusive
tunnel junction, $eV = 2.5\Delta$.} \label{circuit}\vspace{-5mm}
\end{figure}

A convenient interpretation of \Eqs{I002} and \eqref{recurr} in
terms of circuit theory \cite{Bezugly} is given by an infinite
network in the energy space with the period $eV$, graphically
presented in Fig.~\ref{circuit}. The electric current spectral
density\ $J(E)$ consists of partial currents $j_k$, which flow
through the tunnel ``resistors'' $\rho_k$ connected to adjacent
nodes having ``potentials'' $n_k$ and $n_{k-1}$. At $|E|>\Delta$,
the nodes are also attached to the distributed ``equilibrium
source'' $n_F(E)$ through equal resistors $r$. Below we impose the
equilibrium quasiparticle distribution at $|E|>\Delta$, $n(E)=
n_F(E)$, neglecting the effect of small resistors $r$.

The currents flowing between the nodes outside the gap are related
to the thermal current; at $T=0$, these nodes have equal
populations ($n_k=1$ at $E_k<-\Delta$, $n_k=0$ at $E_k>\Delta$),
thus the corresponding partial currents are zero, and the thermal
current vanishes. As a result, only the current $j_0$ flowing
across the gap through the resistor $\rho_0$ survives at $T=0$.

Taking the DOS in the BCS form $N=N_S \equiv \re (E/\xi^R)$, we
see that if any node falls into the gap, the adjacent resistances
turn to infinity, and the current vanishes. For this reason, the
network period must exceed the gap, $eV>2\Delta$, and the
integration in \Eq{I001} is confined to the region $\Delta < E <
eV - \Delta$. This recovers the tunneling model result for the
single-particle tunnel current:\cite{Werthamer} the current
appears above the threshold, $eV = 2\Delta$, having the threshold
value $I_1(2\Delta)= \pi\Delta/2eR $.

To evaluate the subgap current, $eV <2\Delta$, the DOS must be
calculated to next order in the parameter $W$, which requires
solution of the equations for $\hat{g}$ following from
\Eqs{Keldysh} and (\ref{Boundary2}). Using the standard
parametrization $\hat{g}= \sigma_z e^{\sigma_x\theta}$, we obtain
the equation and the boundary condition for the spectral angle
$\theta$,
\begin{align}
&\sinh(\theta - \theta_S) = i \partial_{zz}\theta\sinh\theta_S ,
\;\;\;z = x/\xi_0,
\label{Eqtheta}\\
&\partial_z\theta  +W\sinh\theta (\cosh\theta_1 +
\cosh\theta_{-1})=0\quad (z=-0). \label{Boundtheta}
\end{align}
With exponential accuracy, the solution of \Eq{Eqtheta} for $z<0$
can be approximated by the formula for a semi-infinite wire,
\begin{align}
\tanh\{[\theta(z) - \theta_S]/4\} = \tanh[({\theta_{-0} - \theta_S
})/{4}] \exp(kz),  \label{Soltheta}
\end{align}
where $k^{-1}(E) ={\sqrt{i\sinh\theta_S}}$. Equation
\eqref{Soltheta} describes the decay of perturbations of the
spectral functions at distances $\gtrsim \xi_0$ from the barrier,
where the spectral angle approaches its bulk value $\theta_S=
\arctanh (\Delta/E)$. The boundary value of $\theta$ is to be
found from the equation following from \Eqs{Boundtheta} and
\eqref{Soltheta},
\begin{equation}
2k\sinh [({\theta_S - \theta})/{2}] =W\sinh\theta (\cosh\theta_1 +
\cosh\theta_{-1}). \label{Eqtheta0}
\end{equation}
A direct expansion of $\theta$ with respect to $W$ in
\Eq{Eqtheta0} leads to the following expression for the DOS within
the BCS gap,
\begin{equation}\label{DOSSPT}
N(E)=W({1 - E^2/\Delta^2})^{-5/4} [N_S(E+eV)+N_S(E-eV)].
\end{equation}
The DOS divergencies at $|E|= \Delta, \Delta -eV$ in \Eq{DOSSPT}
are potentially dangerous (cf. Refs.~\onlinecite{MPT}), but they
can be eliminated by improving the perturbation proce\-dure by
solving a set of recurrences in \Eq{Eqtheta0} in the vicinity of
these points.

\begin{figure}[tb]
\epsfxsize=8.5cm\epsffile{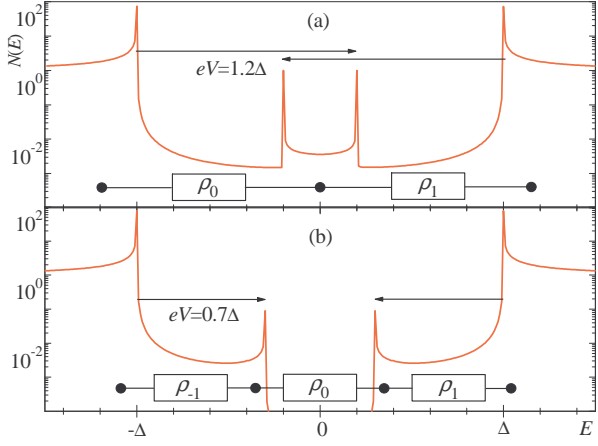}\vspace{-2mm}
\caption{DOS and subgap circuits at the applied voltages
$eV=1.2\Delta$ (a) and $0.7\Delta$ (b), for the tunneling
parameter $W =10^{-3}$.} \label{DOS}\vspace{-4mm}
\end{figure}

As follows from \Eq{DOSSPT}, the tunneling processes transfer the
DOS in the energy space into the BCS gap at the distances $\pm eV$
from the regions $|E|>\Delta$, thus forming an effective spatial
{\em potential well} of the width $\sim \xi_0$ at the tunnel
barrier. At $eV > \Delta$ the BCS gap is entirely filled with the
quasiparticle states with a small local DOS $\sim W$, as shown in
Fig.~\ref{DOS}(a). The appearance of localized states enables the
quasiparticles to overcome the BCS gap at $eV < 2\Delta$ via two
steps involving intermediate Andreev reflection at  energies
$|E|<\Delta$. The population of the intermediate state cannot be
taken to be in equilibrium because the subgap quasiparticles
cannot access the equilibrium electrodes. In the  circuit terms,
the node $k=0$ is disconnected from the equilibrium source, and
the subgap current flows through the two large resistances
$\rho_0,\rho_1 \sim W^{-1}$ (two-particle current), see
Fig.~\ref{DOS}(a). The corresponding partial currents are equal,
$j_0 = j_1 = [{n_F(E_1)-n_F(E_{-1})}]/({\rho_0 + \rho_1})$, and
their contribution to $I(V)$ is confined to the energy region $0<
E < eV-\Delta$ (a similar contribution at $\Delta < E < eV$ comes
from $j_0$ and $j_{-1}$). Thus the two-particle current appears
above the threshold $eV = \Delta$, having the threshold value
$I_2(\Delta) = \pi W\Delta/eR = 2WI_1(2\Delta)$. At $eV= 2\Delta$,
the two-particle current exhibits a sharp peak with the height
$I_2(2\Delta) \approx 2.3 {W^{2/5}\Delta }/{eR}$; at larger
voltages, it approaches a constant value giving rise to the excess
current $I_{\textit{exc}}\approx 6.2 {W^{2/3}\Delta }/{eR}$.

At $eV < \Delta$, a minigap opens in the DOS around the zero
energy [see Fig.~\ref{DOS}(b)], however, since the number of
subgap resistors increases up to three (three-particle current),
the current across the minigap will persist as long as the network
period exceeds the minigap size, $eV> 2(\Delta-eV)$, i.e., at $eV
>2\Delta/3$. The central resistance $\rho_0$ is large,
$\rho_0 \sim W^{-2}$, and dominates the net subgap resistance.
This leads to a smaller charge current with the threshold value
$I_3(2\Delta/3) \approx 2WI_2(\Delta)$. At $eV<2\Delta/3$ the
network period becomes smaller than the minigap, and further
correction to the DOS is required.

Similar results were found for the planar junction
Fig.~\ref{model}(b), using the equation for the functions
$\check{G}_{\pm 0}$ at the top ($+0$) and bottom ($-0$) sides of
the tunnel barrier, $i[\sigma_z E + i\sigma_y
\Delta,\check{G}_{-0}] = 2\Delta W[\check{G}_{-0},
\check{G}_{+0}]$, with the modified tunneling parameter
$W=(3\xi_0^2/4\ell d)D$. This equation is derived by averaging
\Eq{Keldysh} over the thickness of overlapping leads and using
\Eq{Boundary2} (cf. Ref.~\onlinecite{VolkovSIN}). From this
equation we obtain a relation for the spectral angle that does not
significantly differ from \Eq{Eqtheta0},
\begin{equation}
k^2\sinh ({\theta_S - \theta}) =W\sinh\theta  (\cosh\theta_1 +
\cosh\theta_{-1}), \label{Eqtheta1}
\end{equation}
thus giving results which are close to those for the 1D model with
the same magnitude of the parameter $W$.

\begin{figure}[tb]
\epsfxsize=8.5cm\epsffile{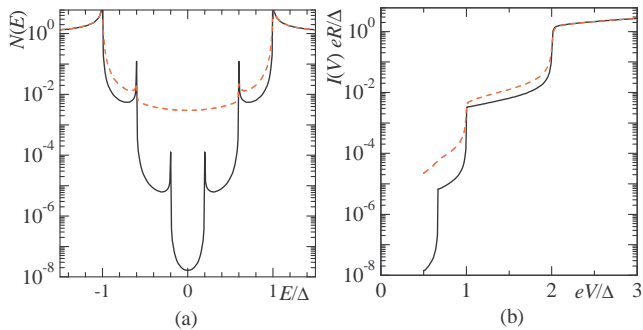}\vspace{-3mm}
\caption{DOS at $eV=0.4\Delta$ (a) and $I$-$V$ characteristics (b)
for the tunneling parameter $W=10^{-3}$ and two values of the
damping parameter: $\gamma =0$ (solid line) and $\gamma=
0.003\Delta$ (dashed line).}  \label{iv}\vspace{-5mm}
\end{figure}

The presented calculation scheme, combining circuit theory
arguments with DOS iteration procedures, suggests an  appealingly
simple explanation for the diffusive SGS: the decreasing applied
voltage results in a shrinking period of the network in
Fig.~\ref{circuit}; hence a stepwise increase of the number of
subgap resistors involved; simultaneously, the number of  DOS
steps, scaled as $W^n$, increases, as shown in Fig.~\ref{iv}(a).
This results in the current staircase with the height of the steps
given by $I_n\sim (2W)^{n-1}I_1$, at $2\Delta/n<eV<2\Delta/(n-1)$.
The quantitative result for the current at arbitrary voltages and
temperatures is
\begin{align}
I(V) &= \int_0^{eV} \frac{dE}{eR} \frac{N_+ + N_-}{\rho_\Delta}
(n_- -n_+) + \int_\Delta^{\infty} \frac{2dE}{eR\rho_1} (n_F -
n_{F1}),\nonumber \\ N_\pm &= \text{Int}\,[(\Delta \mp
E)/eV]+1,\quad n_\pm(E) = n_F(E_{\pm N_\pm}).\label{Inet}
\end{align}
In this equation, the second term represents the thermal current,
the integers $\pm N_\pm$ are the indices of the nodes closest to
the gap edges outside the gap, Int$(x)$ denotes integer part of
$x$, and the quantity $\rho_\Delta(E) = \sum\nolimits_{k= 1-N_-
}^{N_+} \rho_k$ has the meaning of net subgap resistance. The
subgap distribution function reads
\begin{align} \label{nnet}
n(E)=n_+ + (n_- -
n_+)\rho_\Delta^{-1}\sum\nolimits_{k=1}^{N_+}\rho_k.
\end{align}

Equations \eqref{Inet} and \eqref{nnet} are the main technical
results of the paper. The $I$-$V$ characteristic (IVC) of the
planar tunnel junction calculated from \Eqs{Inet} and
\eqref{Eqtheta1} at $T=0$ and shown in Fig.~\ref{iv}(b), was found
to be very close to the result for a ballistic point contact
\cite{Bratus95} with the effective transparency $D_{\textit{eff}}
= 4W= (3\xi_0^2/\ell d)D$. This justifies our statement made in
the introduction, and is the main conclusion of this paper.

In low-transmissive junctions, enhanced subgap current at
$eV<\Delta$ has been observed (see, e.g.,
Ref.~\onlinecite{Gubrun2001}). This anomaly might be due to
many-body interaction effects which introduce a finite lifetime
(damping) of the quasiparticles. The damping effect can be
qualitatively modeled by a small imaginary addition to the energy
in the spectral functions, $E \to E+i\gamma$. This would lead to a
small residual DOS within the BCS gap and cut the DOS staircase at
the level of the order of $\gamma/\Delta$, see Fig.~\ref{iv}(a).
This will result in the smearing of the tunneling SGS and
crossover to a linear IVC at low voltages, $I=2.2(\gamma/\Delta)^2
V/R$, similar to the incoherent MAR regime.\cite{Bezugly} The IVC
calculated from \Eq{Inet} for $\gamma = 0.003\Delta$ and shown in
Fig.~\ref{iv}(b) by a dashed line confirms these considerations.

We conclude our analysis with the estimation of the contribution
of higher harmonics of the functions $\hat g$ and $\hat G^K$ to
the dc current. At $T=0$, the contribution $\delta I$ of the first
harmonics $|m|=1$ (the higher harmonics, $|m|>1$, are smaller,
$\sim W^m$) is
\begin{equation}\label{}
\delta I = \frac{2W}{eR} \int_0^{eV} dE\,\im v
 \im \Bigl(\frac{v}{p} \cosh^2\frac{\chi}{2} + \frac{v}{q}
 \sinh^2\frac{\tilde\chi}{2}\Bigr),
\end{equation}
where $\chi = \theta_{1}+ \theta_{-1}$, $\tilde\chi = \theta_{1}+
\theta_{-1}^\ast$,  $(p,q)^2 = ({\xi_{1}^R +
\xi_{-1}^{R,A}})/{2i\Delta}$, and $v=\sinh\theta$. At $eV <
\Delta$, the energy $E_{-1}$ appears in the subgap region, where
$\theta^*_{-1} = \theta_{-1}+\pi i$ and $\xi_{-1}^A =\xi_{-1}^R$;
for this reason, $\delta I$ turns to zero at $eV < \Delta$,
similar to $I_2$. Numerical calculations show that the
contribution of the first harmonics to the IVC does not exceed
$30\%$. From this we conclude that the adopted quasistatic
approach gives a rather good approximation to a complete solution.

In our treatment, we have neglected inelastic scattering, which
might affect the quasiparticle distribution at subgap energies.
Analysis shows that this effect becomes essential under the
condition $W\tau_\epsilon \Delta \ll 1$, where $\tau_\epsilon$ is
the relaxation time. However, this does not affect the estimate of
the effective scaling factor and only changes the details of the
IVC shape.

In conclusion, we have developed a theory of  subgap charge
transport and subharmonic gap structure in superconducting tunnel
junctions with planar geometry and diffusive thin-film electrodes.
We found that the role of scaling factor in the sub\-har\-mo\-nic
gap structure is played by the effective tunneling transparency
$D_{\textit{eff}}=(3\xi_0^2/\ell d)D$, which may greatly exceed
the bare transparency $D$ of the junction tunnel barrier.

Support from the Swedish grant agencies SSF OXIDE, VR, and KVA is
gratefully acknowledged.

\end{document}